# Title

Edmond Becquerel's colour photographic process involves a Ag/AgCl-based sensitized layer

# Authors list

Victor de Seauve[a, b*], Marie-Angélique Languille[a*], Stéphanie Belin[c], James Ablett[c], Jean-Pascal Rueff[c, d], Christine Andraud[a], Nicolas Menguy[e], Bertrand Lavédrine[a].

[a] Centre de Recherche sur la Conservation (CRC), Muséum national d'Histoire naturelle, CNRS, Ministère de la Culture, 36 rue Geoffroy Saint Hilaire, 75005 Paris, France.
[b] SACRe (EA 7410), Ecole normale supérieure, Université PSL, 75005 Paris, France.
[c] Synchrotron SOLEIL, L'Orme des Merisiers - St. Aubin-BP 48, 91192 Gif s/Yvette, France.
[d] Sorbonne Université, CNRS, Laboratoire de Chimie Physique – Matière et Rayonnement, LCPMR, F-75005 Paris France.
[e] Institut de Minéralogie, de Physique des Matériaux et de Cosmochimie (IMPMC), Sorbonne Université, CNRS UMR 7590, IRD UMR 206, MNHN, Paris, France.

* Marie-Angélique Languille, malanguille@mnhn.fr
* Victor de Seauve, victor.de-seauve@cicrp.fr

# Abstract

Edmond Becquerel invented in 1848 the first colour photographic process. The "photochromatic images" he produced raised several questions, among which the photochromic nature of the sensitized layer. Here, we present the first characterization of the sensitized layer created at the surface of a silver plate or a silver foil according to Becquerel's process, which XIX[th] century scientists called "silver photochloride". It is constituted by silver nanoparticles dispersed in a micrometric silver chloride grains matrix. It is thus similar to the widely studied Ag/AgCl composites, which suggests new synthesis routes for these photocatalysts. The chemical composition of the sensitized layer has been identified by complementary spectroscopies (EDX, XPS, HAXPES and EXAFS), while its morphology has been studied by electron microscopies (SEM and STEM). These techniques involve X-ray or electron beams, which can have an impact on the silver chloride-based sensitized layer; a large part of this article hence introduces a study of the beam effects and of the way of reducing them.

# Introduction

Early photographic processes attract researchers attention for the better understanding of images structures and formations.[1,2] Among these the first colour process, which Edmond Becquerel developed in 1848,[3] raises several questions. This direct positive process requiring no development step was not much employed because of its lack of sensitivity and of the inability to stabilize the coloured images. If the origin of colours of the *photochromatic images* produced by Edmond Becquerel motivated an intense debate among scientists during the second part of the XIX[th] century,[4–7] which we addressed in another paper,[8] the nature of the sensitized layer on which these colours appear was less discussed. Characterizing the sensitized layer is a crucial step in the understanding of the origin of the photochromatic images colours, but it can be complicated by the sensitivity of photographic materials in general, and silver chloride in particular, to usual chemical probes using X-ray or electron beams.

For the development of this process, Becquerel based his researches on previous works concerning silver chloride, notably Herschel's[9] and Hunt's.[10] It is generally admitted, as Becquerel wrote, that the sensitized layer is constituted by "violet [silver] sub-chloride" (all the translations from German and French to English are from the authors), of chemical formula "$Ag^2Cl$" or by a combination of silver chloride and "silver subchloride".[3] Mathew Carey Lea studied the photochromic properties of what he called "red chloride" or "silver photochloride", supposedly constituting the sensitized layers of Edmond Becquerel's photochromatic images, and described it as a combination of both "silver subchloride" and silver chloride.[4] Neither William de Wiveleslie Abney nor Otto Wiener, who both developed hypothesis on the origin of the Becquerel photochromatic images colours,[5,6] questioned the chemical composition of the layers. During the XIX[th] century, only Wilhelm Zenker made the hypothesis that coloured photochromatic images are constituted of regularly spaced "silver dots" (*Silberpünktchen*) in a homogenous silver chloride matrix that would form an interferential pattern.[7]

In addition to be affected by light,[11] the sensitized layer is sensitive to X-ray or electron beams. The main effect caused by X-ray or electron beams to silver chloride-based samples is radiolysis, the chemical decomposition of the material by the ionizing radiation.[12] Sesselman and Chuang hence observed the partial reduction of silver chloride to metallic silver under a 1486.6 eV X-ray beam.[13] Some authors reported that lowering the temperature greatly diminishes this effect.[14–16] The partial reduction of silver chloride to silver nanoparticles has also been observed in a scanning electron microscope (SEM) equipped with a field effect gun (FEG) by Shi *et al.* on a dispersion of synthesized silver chloride nanoparticles.[17] This effect seems to be governed by an accelerating voltage threshold; beyond this, an increase of the accelerating voltage has no effect on the growth rate of the silver nanoparticles, which is typical of the radiolysis effect.[12] In contrast, an increase of the electron flux accelerates the reduction of silver chloride.[17] This effect was also observed by various authors in transmission electron microscopy (TEM).[18,19] Depositing a gold layer on the silver chloride grains[17] or working at liquid nitrogen temperature[12,19] can greatly hamper the radiolysis effect.

The sensitized layer at the basis of Becquerel process appears to be similar to the plasmonic photocatalysts Ag/AgCl and Ag@AgCl. In this type of compounds, a Schottky barrier is created at the junction of the silver nanoparticles, which absorb visible light by surface plasmon resonance, and of the silver chloride, which is a semiconductor. The Schottky barrier allows the formation of holes toward the semiconductor and electrons toward the silver nanoparticles, which can degrade organic pollutants. These photocatalysts are stable and efficient in the visible range.[20,21]

We present here the first characterization of the sensitized layers obtained by the sensitization step of Becquerel process. As a prerequisite, a study of the beam effects occasioned by the analysis using X-ray and electron beam is presented, along with the experimental setups and conditions we used in order to prevent beam damages. Then we present the identification of the main chemical components of the sensitized layer, and that of the traces elements. This has been made possible by the use of various spectroscopies: energy dispersive X-ray spectroscopy (EDX), X-ray photoelectron spectroscopy (XPS), hard X-ray photoelectron spectroscopy (HAXPES) on the GALAXIES beamline of the SOLEIL synchrotron[22] and X-Ray absorption spectroscopy (XAS) for extended X-ray absorption fine structure (EXAFS) analysis on the ROCK beamline of the SOLEIL synchrotron.[23] And lastly the morphology of the sensitized layer and the chemical elements spatial distribution, studied by scanning transmission electron microscopy coupled with EDX (STEM-EDX), are discussed.

## Experimental

"Real samples". Edmond Becquerel's process of sensitization has been replicated in the lab and the exposition to visible light of sensitized samples has been published elsewhere.[11] In the historical process, a sensitive layer is produced onto a silver coated brass plate, as the ones used for daguerreotypes. We also prepared purer samples by evaporation of silver on a silicon wafer. The silver plate is sensitized either by soaking it in a copper chloride solution ("immersion samples"), either by applying a voltage between the plate and a platinum grid, both of them being soaked in hydrochloric acid ("electrochemical samples"). The sample is then rinsed, dried, and heated. The sensitive layer has a pinkish hue and the colours appear directly onto it as it is exposed to visible light. The samples are exposed to 8 narrow-bandwidth light spots of varying peak wavelengths (402 to 677 nm) and irradiance (0.8 to 2.5 W m$^{-2}$). An optimal radiant exposure to visible light of 10 kJ m$^{-2}$ for the preparation of coloured samples has been estimated. These samples are close to Becquerel's photochromatic images and are thus called here *real samples*.

"Model samples". On the other hand, *model samples* were designed for analytical purposes, in order to analyse in transmission the sensitized layer, and not the underlying unsensitised silver support. In order to isolate this sensitized layer, micrometric thick silver foils were integrally sensitized, resulting in a few microns thick sensitized layers. They were exposed in the same conditions as real samples and at the same radiant exposure.

TEM samples preparation. TEM samples have been prepared from model samples by ultramicrotomy. TEM cross section observations of such samples were consistent with top view observations of model samples in a SEM. Note that the morphology of focused ion beam (FIB) prepared samples was significantly modified by the ion beam (cf. figure S-1 in the SI section), which is why this technique was ruled out. Ultramicrotomy, which has already been used for cutting photographic materials,[24] was then preferred to the FIB preparation. The sections were deposited on carbon coated TEM grids. A carbon deposit was made, sandwiching the sections in carbon in order to protect the samples from the electron beam (see below). The thickness of the sections has been estimated to 70 nm by the log-ratio technique,[25] assuming the sensitive layer is only constituted by silver chloride. During all the preparation, the sample was protected from light with black paint and by working in the dark as much as possible. The irradiance at the sample position during the various steps of the preparation has been determined then the radiant exposure of the sample to ambient light has been estimated. It is the same for all samples, around 2 kJ m$^{-2}$ of polychromatic light without UV. Given the obstacles to the preparation of TEM samples in the dark, this exposition, which represents 20 % of the radiant exposure required to colour the samples, is a great achievement.

Experimental procedures. Comprehensive details about the analytical methods and procedures can be found in the SI section.[26,27]

# Results and discussion

## Beam damage regulation

Dose calculation. When using an X-ray beam or an electron beam, the doses were calculated in order to monitor the samples changes during the analysis. The dose, in gray (1 Gy = 1 J kg$^{-1}$), is the amount of energy deposited by the beam on a sample, per unit of mass. It can be calculated for an X-ray beam[28] as well as for an electron beam.[12,29] In the case of X-ray experiments (XAS, HAXPES), the dose is calculated using the photon flux, the exposure time, the photon energy, the beam surface and the incidence angle α, the sample thickness, the assumption it is constituted only by silver chloride, and the X-ray transmittance of the sample. The latter is measured in XAS experiments, and calculated from the sample thickness and the photon energy[30] with the CXRO website[31]. The dose formula for X-ray experiments is given by equation S-1 in the supporting information (SI) section.

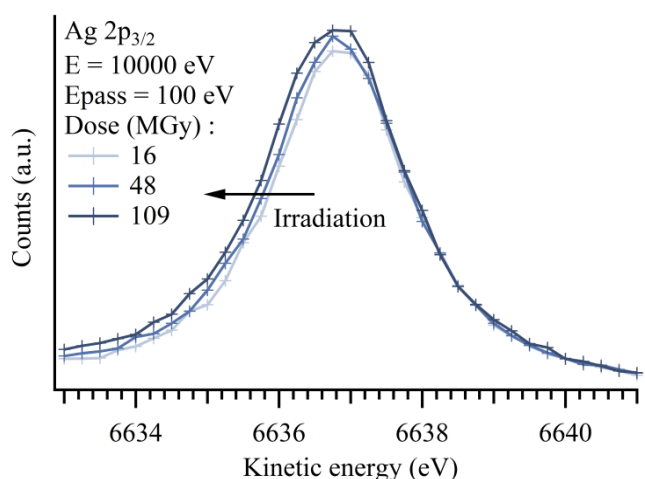

Figure 1. Successive HAXPES spectra in the energy range of Ag 2p$_{3/2}$ electrons of a blue sample sensitized by immersion (silver deposit on a silicon wafer).

For STEM experiments, the dose is calculated using the beam intensity and the electronic charge, the exposure time, the sample thickness, the assumption it is constituted only by silver chloride, the stopping power of silver chloride at the electron energy provided by Berger *et al.*,[32] and the scanned area. The latter is determined by multiplying the number of lines of pixels of an image by the area of a line scan, which is equal to the length of the image multiplied by the diameter of the beam at the sample position. The dose formula for STEM imaging is given by equation S-2 in the SI. For SEM experiments, the dose is calculated using the beam intensity and the electronic charge, the exposure time, the electron energy, the assumption the sample is constituted only by silver chloride and the scanned area, which is calculated the same way as in the STEM. The thickness which absorbs the electrons is determined by Monte-Carlo simulations with the CASINO software.[33] The dose formula for SEM imaging is given by equation S-3 in the SI.

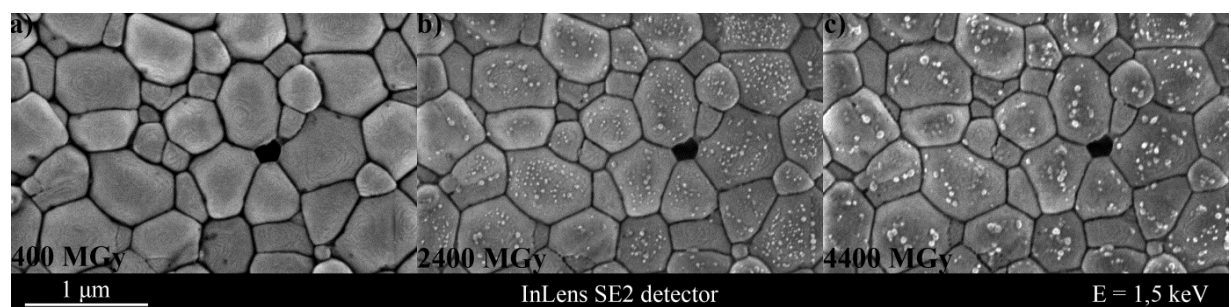

Figure 2. Successive secondary electron images of the same area of a blue sample sensitized by immersion (silver deposit on a silicon wafer). The brightness and contrast of the second and third images have been corrected so that the greyscale dynamic of the three images are comparable.

Radiolysis damage. During the HAXPES analysis of a blue sample at doses as high as a hundred MGy, the Ag $2p_{3/2}$ peak shifted toward lower kinetic energies (figure 1). This is coherent with the reduction of the silver specie to metallic silver. The increase of the Ag $2p_{3/2}$ peak can be attributed to the desorption of the chlorinated specie during X-ray irradiation.[13] Imaging samples in a SEM at doses as high as a few hundred MGy results in the formation and growth of nanoparticles, which appear brighter than the grains constituting the layer (figure 2). The same observation for comparable doses, has been made in STEM, while the accelerating voltage and the electronic dose rate are quite different (200 keV and 15 e$^-$ Å$^{-2}$ s$^{-1}$ for the STEM; 1.5 keV and 0.14 e$^-$ Å$^{-2}$ s$^{-1}$ for the SEM); therefore, in both cases the images have been obtained at accelerating voltage and electron dose rates higher than the threshold values characteristics of the radiolysis effect; in this case, the damages are proportional to the dose.[12,29]

Shi *et al.* reported the absence of radiolysis damage on silver chloride particles coated by a 10 nm gold layer in a SEM.[17] We chose to protect the samples by a carbon deposit, compatible with elemental analysis. No radiolysis damage is observed in the SEM on carbon coated samples. In addition to this, working at liquid nitrogen temperature in the STEM prevents the radiolysis effect, at least until doses as high as a few million MGy.

The radiolysis damage has been reduced in HAXPES by working at 200 K, reducing the dose rate and by continuously moving the sample under the beam, in order to less irradiate the same area of a sample. The latter prevents us from calculating the dose. The damage being reduced, but not avoided, comparisons between HAXPES spectra have been made in the same analysis conditions: on a fresh area of the sample, or at the same dose.

Finally, working during the single bunch mode of the synchrotron machine and using the ROCK beamline Quick-XAS monochromator allowed us to acquire spectra at very low doses. From a few 0.1 MGy to a few MGy, no modifications of the XAS spectra have been detected, possibly because these relatively low doses did not have effects on the samples, or because the lower sensitivity of this technique did not allow the detection of radiolysis damage.

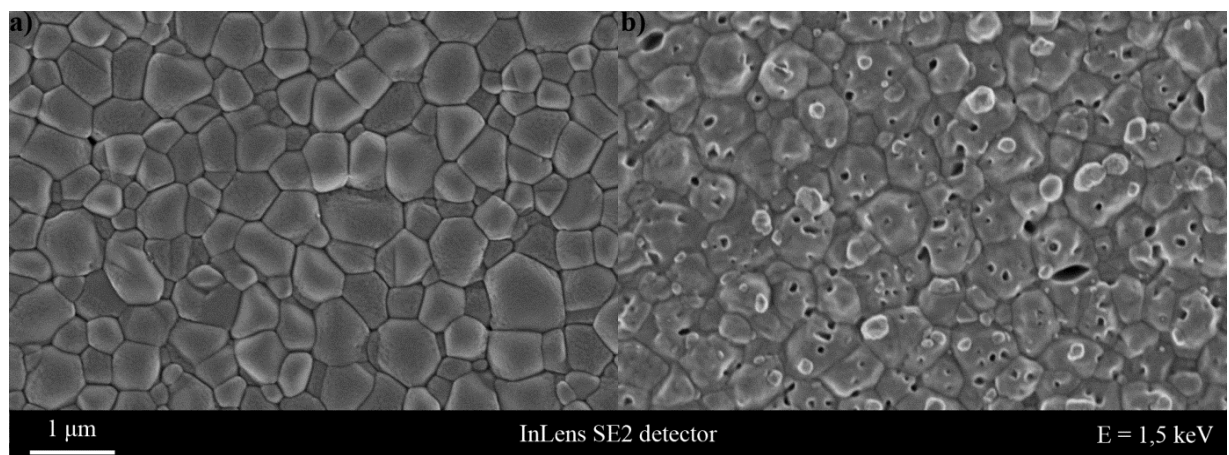

Figure 3. Secondary electron images of a fresh area (a) and an irradiated area (b, a few hundred MGy) of a blue sample sensitized by immersion (silver deposit on a silicon wafer).

<u>Knock-on damage.</u> During extended analysis of radiolysis-protected samples in a STEM, for doses as high as a few hundred thousand MGy, roughness and porosity were created on the sample. This is attributable to knock on damages. This phenomenon usually exhibits an accelerating voltage threshold effect.[12] In the microscope that was used, the voltage is not tuneable; therefore, the only way of avoiding this type of damage is to acquire the data quickly in order to lower the dose accumulated on the sample.

This phenomenon also took place during HAXPES experiments on the static sample. Figure 3 shows SEM images of a fresh area of the sample (figure 3a) and of an area that has been analysed at doses as high as a hundred MGy (figure 3b). Surface roughness and porosity, with a few hundred nanometres wide holes, were created by the X-ray beam during the analysis. This was completely avoided by working at 200 K, reducing the dose rate and by continuously moving the sample under the beam. No such damages were observed after XAS experiments in the conditions described above.

## Characterization results

<u>Bulk composition.</u> EDX analyses conducted in the SEM on real and model samples, electrochemically and immersion-sensitized showed only the presence of silver and chlorine; quantifications by the φ(ρz) method showed an Ag/Cl ratio close to 1 (cf. figure S-2 in the SI section). XRD analysis made on a model sample before and after the sensitization showed the presence of crystallized AgCl with a (1 1 1) preferred orientation. Lastly, the Fourier transform (FT) of the EXAFS signals of a sensitized sample and that of a silver chloride standard (figure 4) have been fitted (see Figure S-3 for the spectra and Table S-1 for the results of the fits) using Hull and Keen's crystallographic parameters.[34] The main peak around 2.2 Å corresponds to the first coordination shell, which is constituted by chlorine atoms, whereas the second silver atoms coordination shells is less noticeable, between 3 and 4 Å. These FTs are very close: the silver chloride contained in the bulk of sensitized samples has the same fine structure as a standard AgCl.

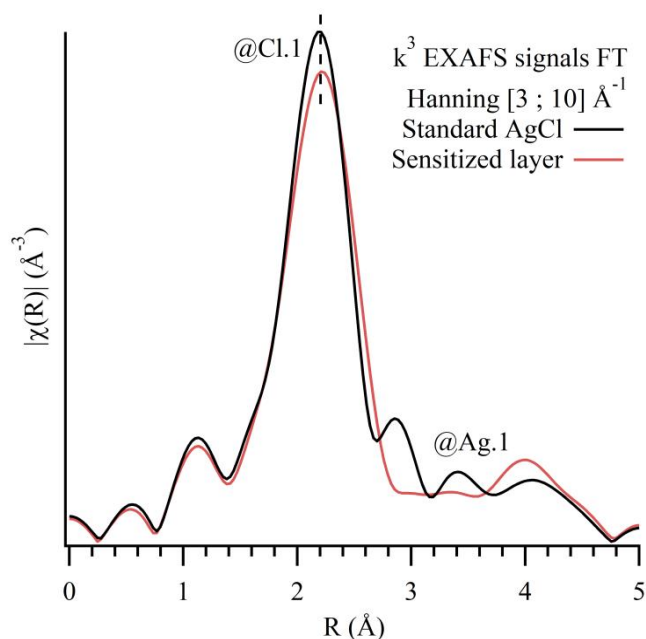

Figure 4. $k^3$ FTs of the EXAFS signals, with a [3 ; 10] Å$^{-1}$ Hanning apodisation window of a sensitized layer (36000 averaged spectra, 5h acquisition time) and a standard AgCl pellet (2300 averaged spectra, 20 min acquisition time).

Surface composition. XPS and HAXPES analyses of real samples showed the presence of the following traces elements: C, O, S, N, Na. These are classically present on samples that are not prepared under vacuum. Only immersion sensitized samples showed traces of Cu. This can be explained by the presence of copper in the immersion sensitization solution that may have not been rinsed properly. The 9.9 keV photoelectrons produced by the HAXPES scanning of the valence band of sensitized samples (figure 5) have an inelastic mean free path in silver of 8.69 nm.[35] The thickness that is analysed in XPS is usually estimated to be 3 times the inelastic mean free path of the photoelectrons; it is reasonable to assume that, if the sensitized samples are only constituted by silver chloride, the thickness that is analysed is of the order of 20 nm. Comparing our data with Mason's results on silver chloride (table 1), we then deduce that the extreme surface of the sensitized layer is constituted by silver chloride.

Table 1. Positions in energy of the features of the valence bands of AgCl, as measured by Mason,[36] and of the sensitized layer.

| Features | AgCl energy (eV) | Sensitized layer energy (eV) |
|---|---|---|
| Top | 1.55 | 1.8 |
| Shoulder | 3.8 | 4.2 |
| Maximum | 4.5 | 4.8 |
| Peak | 5.7 | 6.1 |

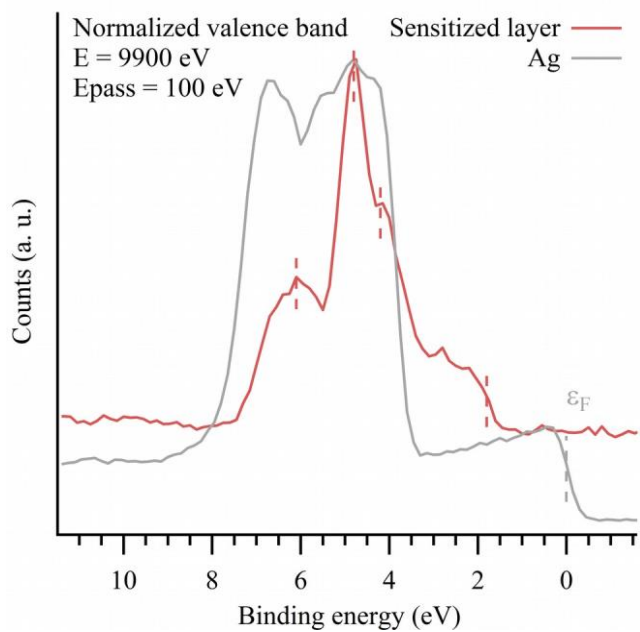

Figure 5. HAXPES spectra in the valence band energy range of a sensitized layer (pink) and of a standard silver deposit (grey). The zero of binding energy was determined from the Fermi level of silver.

Surface morphology. Figure 6 shows SEM images in top view of real samples sensitized by immersion (figure 6a) and by electrochemistry (figure 6b). On the electrochemically sensitized sample, the grains are facetted in growth plans, with an average size of 0.82 µm, whereas they are rounded, with less definite edges, with an average size of 0.72 µm on the immersion sensitized sample (cf. figure S-4 in the SI section for the size distributions). These facies are coherent with what Song *et al.* observed on a silver sheet immersed in an iron chloride (II) solution.[37] It is interesting to point out that not all electrochemically prepared samples are facetted in growth plans.

Visible absorption. Figure 7 shows the average UV-visible absorptance spectrum of 14 sensitized model samples and that of a standard silver chloride pellet. The latter displays a long absorption tail in the visible region, a steep edge in the violet region (404 nm), and a second edge in the ultraviolet region (255 nm), which is consistent with previous studies.[38,39] The sensitized layers spectrum is comparable to that of AgCl in the ultraviolet region, but shows a broad absorption band in the visible region, centred on 500 nm. According to several authors having studied the photolysis of silver chloride, this band is characteristic of the presence of silver nanoparticles.[40–42]

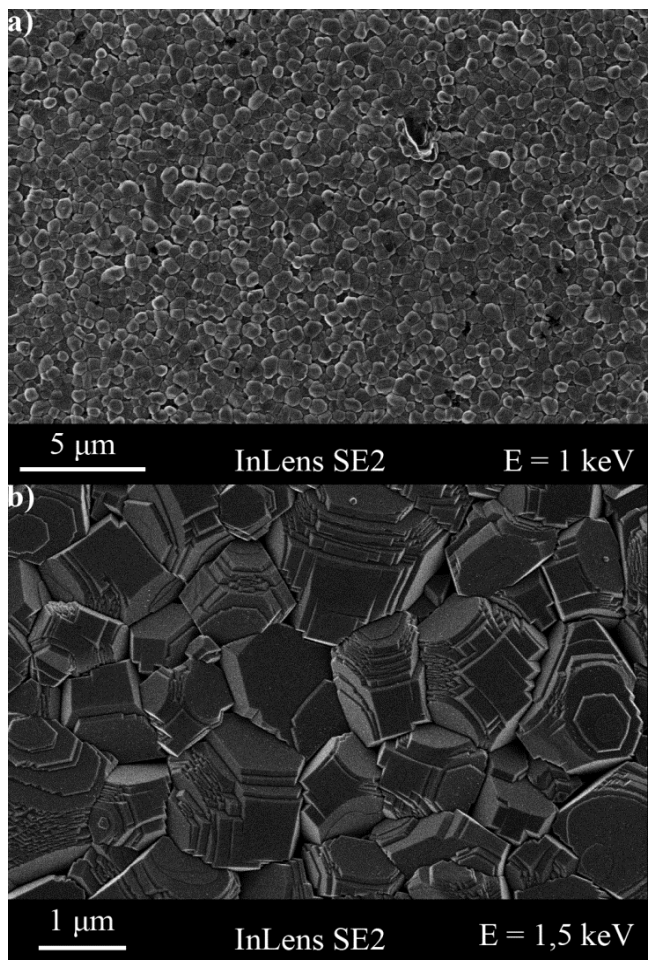

Figure 6. Secondary electron images of real samples sensitized by immersion (a) and by electrochemistry (b).

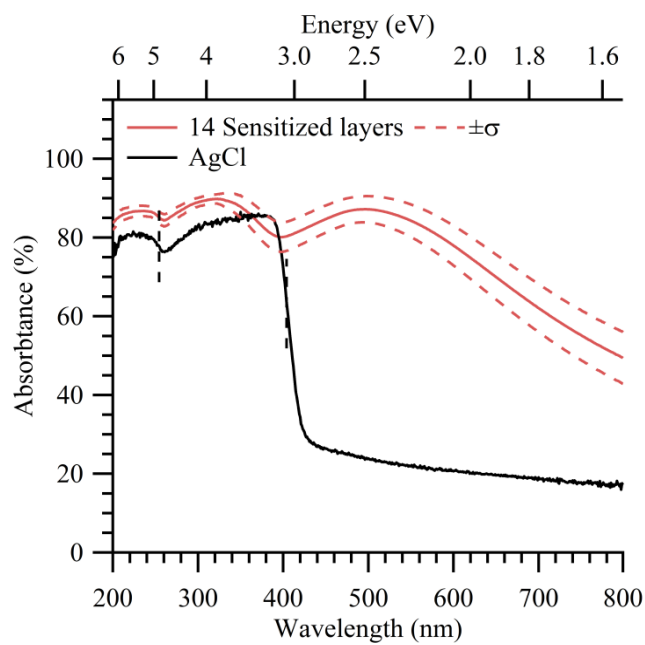

Figure 7. Average UV-visible absorptance spectra of 14 model samples compared with a standard AgCl pellet spectrum.

Bulk morphology. Figures 8 and 9 show STEM images using the chemical contrast providing high angle annular dark field (HAADF) detector in cross-sectional view at low (figure 8) and high magnification (figure 9). Figure 8 shows the whole sensitized layer sample; it is constituted by micrometric grains, which is coherent with the plane view of figure 6b. The average thickness of the layers we studied is 2.7 µm, which is coherent with the predictions, assuming that a 1 µm silver foil is entirely transformed in silver chloride (cf. equation S-4 in the SI section). This is also coherent with previous calculations based on charge transfer calculations for real samples which give a 3.7 µm thickness, and with the optimal 1.7 µm thickness which Becquerel calculated.[11] In the same way as the morphology in plane view, slight variations in the bulk morphologies were observed. This can be explained by the variability in the preparation of the samples, and particularly during the sensitization – variability in the distance between the silver foils at the platinum grid – and during the heating – distance between the heat gun and the sensitized layers. These various morphologies explain the dispersion in the reflectance spectra of the sensitized layers that were previously observed[11] and in the absorptance spectra (figure 7).

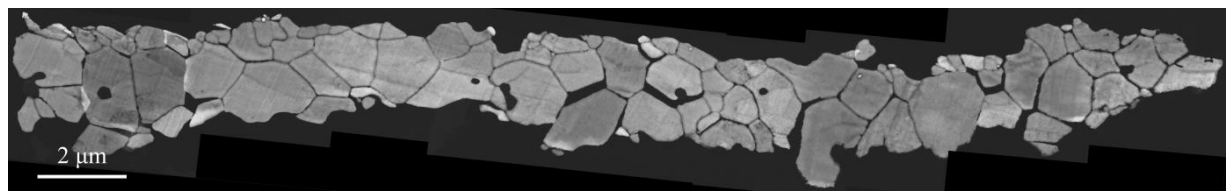

Figure 8. STEM-HAADF stitched images of a blue sample.

Presence of nanoparticles. Figure 9 shows the presence of nanoparticles that appear brighter that the grains, which in this imaging mode is consistent with a higher average atomic number. Most of these nanoparticles are located inside the grains, and some at grain edges; very few are located at grain boundaries. The average size of grain edges particles is 52 nm, that of embedded nanoparticles is 27 nm (cf. figure S-5 in the SI section for size distributions). Moreover, there are more embedded nanoparticles than grain-edges nanoparticles. Figure 10 shows a STEM-EDX map of a grain edge nanoparticle. It appears that it is only constituted by silver, whereas the grains are constituted by silver and chlorine. High resolution TEM (HRTEM) images of a grain and of a nanoparticle showed interplanar spacing corresponding to AgCl (2 2 0) and AgCl (2 0 0) in the grain and Ag (1 1 1) in the nanoparticles (cf. Figure S-6 in the SI section). Becquerel[3] and Carey Lea[4] observed that the sensitized layer – or the "silver photochloride" – contains more equivalents of silver than of chlorine, which lead them to identify it as a mixture of silver chloride AgCl and silver subchloride "$Ag^2Cl$".[11] Their initial observation conformed to the identification of the sensitized layer as silver nanoparticles dispersed in a silver chloride matrix. This finding makes the sensitized layers comparable to the various Ag@AgCl and Ag-AgCl plasmonic photocatalysts that have been developed and widely studied over the last 10 years.[18,20,43,44]

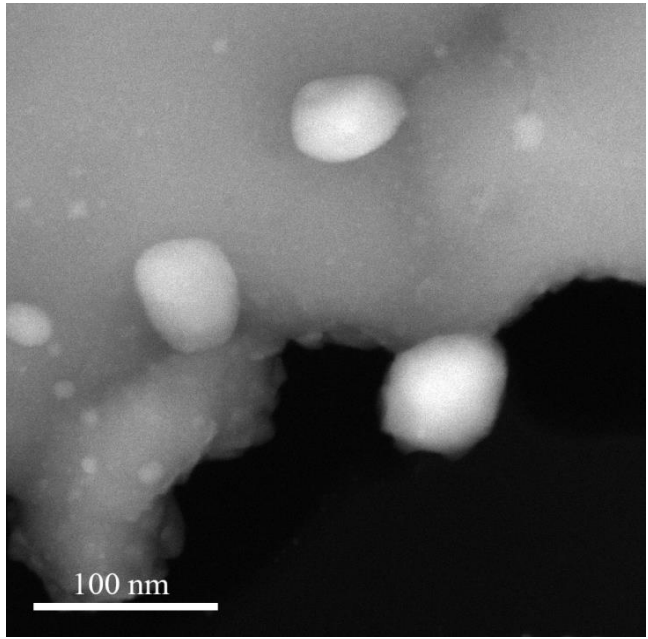

Figure 9. STEM-HAADF image of a detail of a red sample.

<u>Interpretation of XAS, HAXPES and UV-visible results.</u> These last observations confirm that the sensitized layers are constituted by silver nanoparticles dispersed in a micrometric silver chloride grains matrix. Song *et al.* also noticed the presence of silver nanoparticles decorating the silver chloride grains created by immersing a silver sheet immersed in an iron chloride (II).[37] The silver nanoparticles occupy less than 1 % of the volume of the layers; this is too few to be detected by XAS, which is why only silver chloride was detected by this method. Furthermore, the silver nanoparticles are usually not located at the recto or at the verso of the layers, which explains why only silver chloride was detected at their surface by HAXPES. Moreover, the average size of the silver nanoparticles is compatible with Moser *et al.*[42] and Rohloff's calculations,[45] which attribute the broad absorption band in the visible region to silver nanoparticles of various sizes and shapes, which size distribution would be centred on 60 nm.

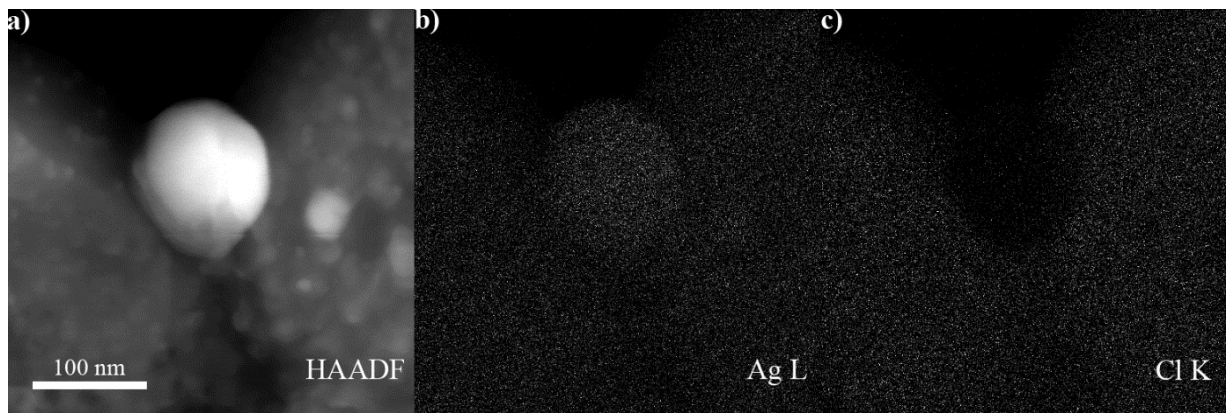

Figure 10. STEM-EDX maps of a silver nanoparticle found in a blue sample.

# Conclusions

Edmond Becquerel's process of sensitizing silver plates or foils was replicated in the lab in order to create real and model samples, the former being merely similar to Becquerel's one and the latter being only constituted by the sensitized layer. Prior to the identification of the sensitized layer, a study of the damages caused by the X-ray and electron beams was conducted: the energies deposited on the samples during the analysis, the doses, were calculated and the beam damages were monitored throughout the acquisitions. Parameters adjustments were made in order to control the beam effects, including reducing the temperature and the dose rates. In particular, a thin carbon film deposited onto the silver chloride-based samples prevented the radiolysis damages that were otherwise observed in the electron microscopes. The elemental composition of the sensitized layer has been determined by SEM-EDX, XPS and HAXPES. The chemical species of the layer have been identified by HAXPES, X-ray diffraction (XRD) and their fine structures have been studied by EXAFS. Silver chloride, the major component, does not explain the absorption properties of the sensitized layers in the visible range. The presence of silver nanoparticles, which absorb in the visible range, was evidenced from cross-sectional observations of the layers in a TEM. The low concentration of metallic silver and its location in the layers explain why it was undetected by XAS and HAXPES.

The layers created during the sensitization step of the photographic process are constituted by micrometric silver chloride grains and silver nanoparticles in the size range [10 – 150] nm. The former constitutes the rough and porous sensitized layer; the latter can be located inside the grain or at grain edges, rarely at grain boundaries. This allows us to identify what was called "silver photochloride" during the XIX[th] century as being constituted by silver nanoparticles dispersed in a silver chloride matrix, which makes it comparable to the Ag/AgCl composites which have recently drawn a lot of attention in the plasmonic photocatalysis field.[21] This suggest new synthesis routes for these composite materials, susceptible to be of interest, given that it does not involve organic solvent,[43] nor exposition to UV light[18,44] and that the chemicals that are used are quite cheap compared to other approaches.[20]

# Acknowledgements

The authors would like to thank the SACRe doctoral program (PSL University) for the financial support. This work was supported by the Observatoire des Patrimoines de Sorbonne Universités (OPUS). Thanks to the beamtime allocation n° 20150273, HAXPES experiments were performed on the GALAXIES beamline at SOLEIL Synchrotron, France. The authors wish to acknowledge the award of beamtime n° 20160485 on the ROCK beamline at Synchrotron SOLEIL which was supported by a public grant overseen by the French National Research Agency (ANR) as a part of the "Investissements d'Avenir" program ref: ANR-10-EQPX-45. The authors would like to thank Christophe Méthivier from Laboratoire de Réactivité des Solides (CNRS) with who the XPS analysis were performed.

# Supplementary Information

**Equation S-1.** Dose formula for X-ray experiments.

$$Dose(\text{Gy}) = \frac{Photon flux(\text{ph.s}^{-1}) \times E_{photon}(\text{J}) \times \sin\alpha \times Exposure time(\text{s}) \times (1 - Transmittance)}{Beam surface(\text{m}^2) \times Thickness(\text{m}) \times d_{AgCl}(kg.\text{m}^{-3})}$$

**Equation S-2.** Dose formula for STEM experiments.

$$Dose(\text{Gy}) = \frac{Beam intensity(\text{A}) \times Exposure time(\text{s}) \times Total Stopping Power_{AgCl,E}(\text{J.m}^2.\text{kg}^{-1})}{Electron charge(\text{C}) \times Number of lines \times image length(\text{m}) \times Beam diameter(\text{m})}$$

**Equation S-3.** Dose formula for SEM experiments.

$$Dose(\text{Gy}) = \frac{Beam intensity(\text{A}) \times Exposure time(\text{s}) \times E_{elecctron}(\text{J})}{Electron charge(\text{C}) \times Scanned area(\text{m}^2) \times Absorbing thickness(\text{m}) \times d_{AgCl}(kg.\text{m}^{-3})}$$

**Equation S-4.** Formula for the calculation of the sensitized layer thickness, assuming all the silver is transformed in silver chloride. M represent a molar mass, ρ a density.

$$Sensitized layer thickness = Ag layer thickness \times \frac{\rho_{Ag} \times M_{AgCl}}{\rho_{AgCl} \times M_{Ag}}$$

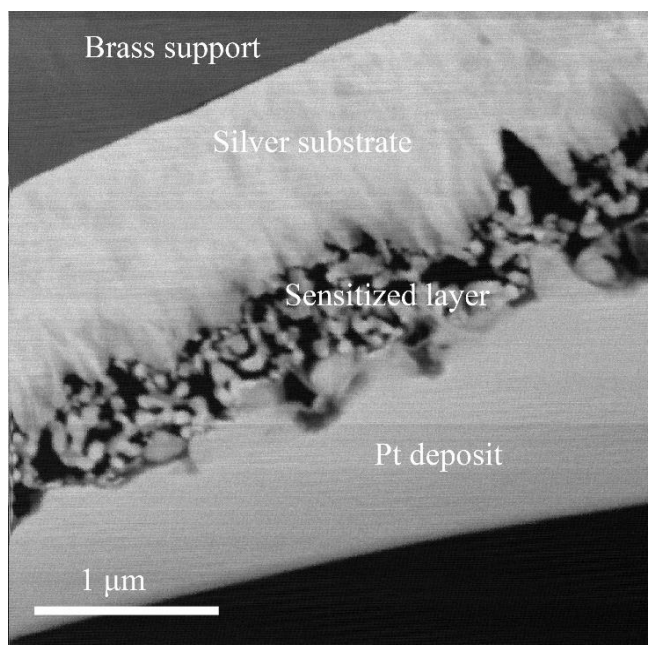

**Figure S-1.** STEM-HAADF image of a FIB-prepared sensitized real sample. The sensitized layer appears rough and porous; it has probably been damaged by the ion beam.

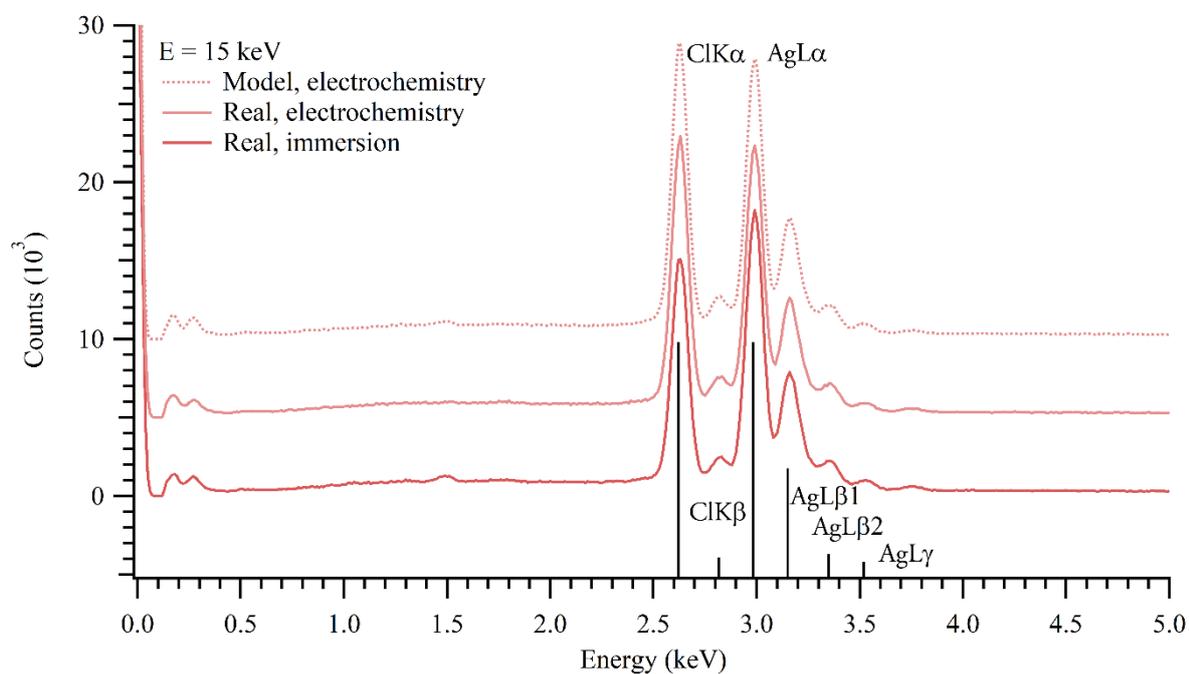

**Figure S-2.** EDX spectra of model and real samples sensitized by electrochemistry and immersion. The Ag/Cl ratio calculated by the φ(ρz) method are 0.93 for the electrochemically sensitized samples (model and real) and 1.17 for the immersion-sensitized real sample.

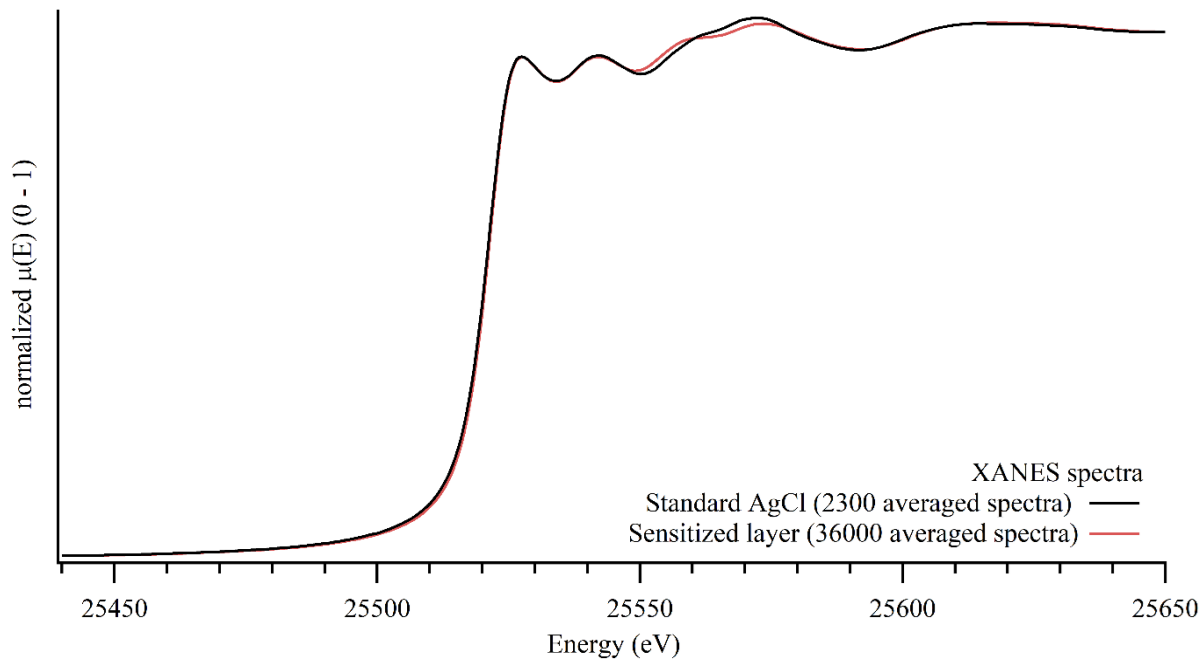

**Figure S-3.** XAS spectra of a sensitized layer and a standard AgCl pellet.

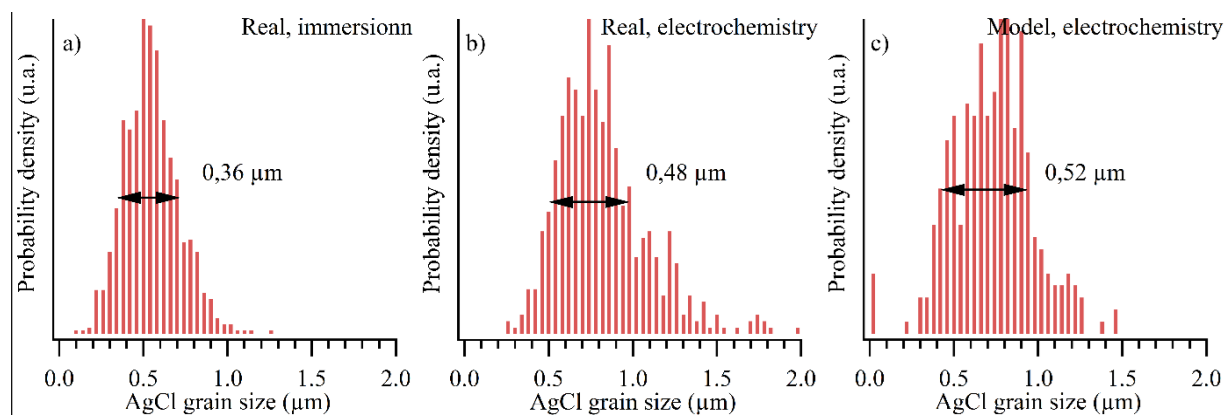

**Figure S-4.** Silver chloride grain size distributions, determined from plane view images obtained in the SEM, for real samples sensitized by immersion (a), by electrochemistry (b) and for model samples (c).

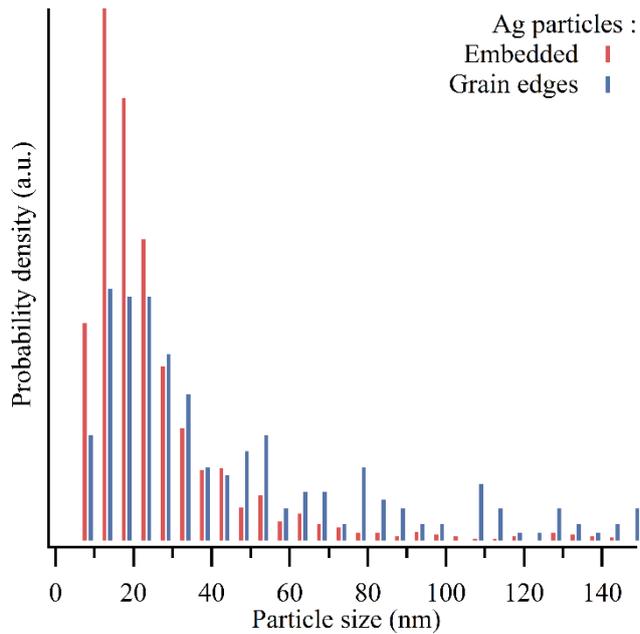

**Figure S-5.** Silver nanoparticles size distribution, according to their localization. 1451 embedded nanoparticles and 268 grain-edge nanoparticles were measured.

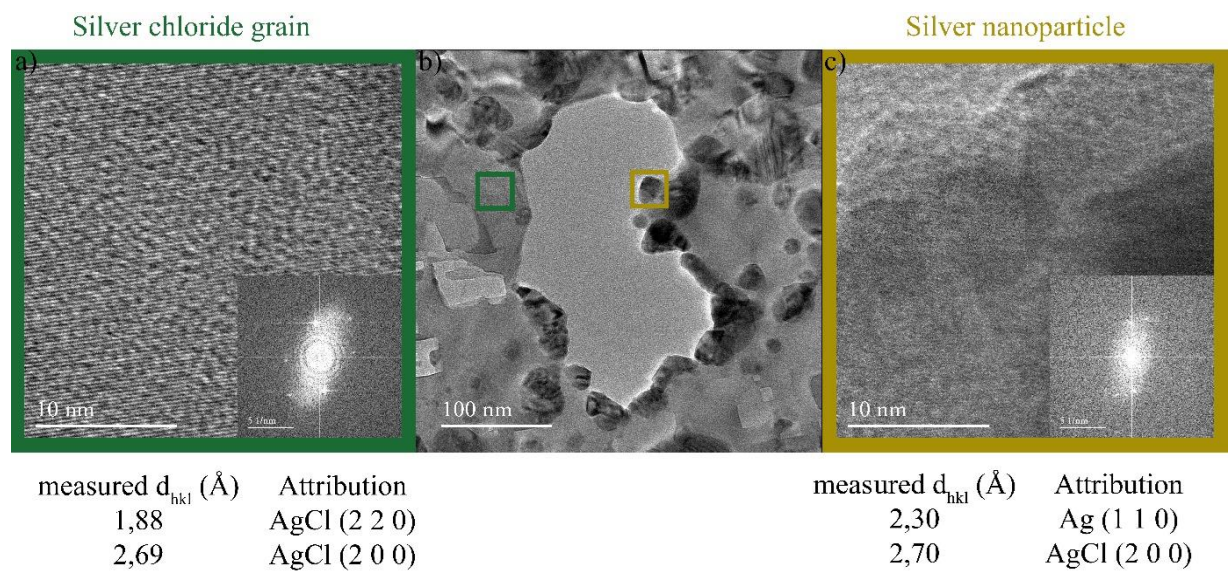

| measured $d_{hkl}$ (Å) | Attribution |
|---|---|
| 1,88 | AgCl (2 2 0) |
| 2,69 | AgCl (2 0 0) |

| measured $d_{hkl}$ (Å) | Attribution |
|---|---|
| 2,30 | Ag (1 1 0) |
| 2,70 | AgCl (2 0 0) |

**Figure S-6.** HRTEM images of a model sample. b) Global view. a) HRTEM image of the area circled in green, which corresponds to a silver chloride grain. Inset: FT of the HRTEM image. Table: $d_{hkl}$ measured on the FT and the plane attributions. c) HRTEM image of the area circled in yellow, which corresponds to a silver nanoparticle. Inset: FT of the HRTEM image. Table: $d_{hkl}$ measured on the FT and the plane attributions.

|  | R-factor | Sphere number | Crystallographical parameters | | Fits results | | | |
|---|---|---|---|---|---|---|---|---|
|  |  |  | Coordinence | Distance (Å) | $\Delta E_0$ (eV) | Coordinence | Distance (Å) | $\sigma^2$ (Å²) |
| Standard AgCl | 0.0673 | 1 | 6 | 2.773 | -2.28 | 2.163 | -0.088 | 0.0127 |
|  |  | 2 | 12 | 3.922 |  | 3.059 | -0.124 | 0.0305 |
| Sensitized sample | 0.0251 | 1 | 6 | 2.773 | -0.257 | 2.163 | -0.0682 | 0.0171 |
|  |  | 2 | 12 | 3.922 |  | 3.059 | -0.0964 | 0.0343 |

**Table S-1.** Results of the fits of the standard AgCl pellet and of the sensitized sample between 1 and 5 Å (Hanning apodisation window [3; 10] Å$^{-1}$) including the first two coordination shells of AgCl. The $S_0^2$ factor have been set to 0.78, according to the first fit made on the TF of the EXAFS signal of the sensitized samples, which has been acquired with a higher statistic.

**Experimental section.**

Materials. Real samples are made from 30 µm-silver coated 20 × 20 mm² brass platelets. These were ordered from the silversmith François Cadoret, 172-174 rue de Charonne – 75011 Paris, France (no longer in business). They are ¼ µm-diamond polished and cleaned in ethanol before the sensitization. Only silver was detected by XRF on the gilded side. The brass support contains traces of Cr and Co. Model samples are made from 1 µm thick silver foils, ordered from the silversmith Mario Berta Batilloro, Cannaregio 5182 – 30121 Venezia, Italia. Silver and traces of Fe, Cu, Zn and Cr have been detected by XRF in these foils.

SEM-FEG. SEM samples are protected with a carbon deposit made with a Leica EM SCD500 evaporator at 5×10$^{-4}$ mbar. The microscope is a SEM-FEG Zeiss Ultra 55 working at 1.5 kV, with a 10 µm aperture at a working distance ranging from 1 to 2 mm. The "high current" mode is not activated. 1024 × 768 pixels images are taken with the following parameters: "scan speed" 9 and "line average" 3 (1 min/image). The specific detection of SE1 is made with an in-column detector called "In Lens" on this type of microscope. This obliges to work at a low energy and offers a better spatial resolution, but lowers the depth of field. For EDX analysis with the SDD Bruker X-flash 4010 10 mm² detector, the microscope is operated in the following conditions: 15 kV, 120 µm aperture, 7.5 mm working distance, "high current", shaping time 90 kcps, energy range 40 keV, Cu Kα FWHM 146 eV. Spectra were acquired in 80 s. The approximated beam diameter onto the sample is 3 nm. The Optiprobe software allows the measurement of the beam intensity, which is 17 pA in the imaging conditions and 2798 pA in the EDX analysis conditions.

Size measurements of grains have been made with the Fiji software on 880 grains for real immersion-sensitized samples, 577 grains for real electrochemically sensitized samples and 326 grains on electrochemically sensitized model samples.

TEM-FEG. TEM samples were embedded in LR white resin, that polymerised at 50°C and a few 0.1 bar during 36 h. A Leica UC7 ultramicrotome was used, together with a Cryotrim diamond knife for the trimming and an Ultra 35 diamond knife for the cutting of ultrathin sections. The sections are cut with the layer sample parallel to the knife; the recto and the verso of the layers are identified before the embedding and at the end of the preparation, by comparison with the position of the sample in the resin section. 5 nm carbon deposits are made with a Leica ACE600 evaporator in the E-beam mode at $1\times10^{-5}$ mbar, at around 0.2 nm s$^{-1}$.

The microscope is a JEOL 2100F equipped with a Schottky FEG working at 200 kV, a JEOL STEM detector, a Gatan US4000 TEM camera, a JEOL Si(Li) 30 mm² EDX detector (138 eV resolution, bias: 700 V), and a cryo sample holder. The 50 µm aperture was used, with a 0.2 nm spot size for STEM-HAADF imaging. The 1024 × 1024 pixels images are acquired in 33.5 s. The same aperture was used for EDX analysis, with a 0.7 nm spot size. 512 × 512 pixels maps are acquired in ten minutes in the "T4" mode, favouring the energetic resolution over a high saturation limit. Lastly, the same aperture was used for HRTEM imaging, in the following conditions of brightness and beam convergence angle: "spot 1 / α3" for images at magnifications lower than 100 k and "spot 1 / α1" for higher magnifications. 4096 × 4096 images (binning 1) are acquired in 5 s. The phosphorescent screen picoamperemeter was used in order to determine the beam current, which is 57 pA for STEM imaging, 650 pA for EDX analysis and 1100 pA for TEM imaging.

Size measurements of nanoparticles have been made with the Fiji software on 1451 particles for embedded nanoparticles and 268 grain-edge nanoparticles. Thickness measurements have been made on 11 layers.

HAXPES. HAXPES experiments were performed at the GALAXIES beamline at the SOLEIL synchrotron. The X-ray beam provided by an undulator is monochromatised by a double-crystal Si(111) monochromator (DCM) and focused onto the sample position by a toroidal mirror located downstream. Incident energies of 10 keV or 9.9 keV X-ray were used. The third harmonic of the DCM was used in order to lower the photon flux and increase the energetic resolution. In addition, a 5.8 % transmitting filter was used to further attenuate the beam. The photon flux in these conditions was measured at $7\times10^9$ ph s$^{-1}$. The resolution of the monochromator at this energy is around 0.3 eV. The beam size was $20 \times 80$ µm² and the angle between the beam and the sample surface was 20°. The sample was cooled down at 200 K. The beamline is equipped with a EW4000 VGScienta hemispherical analyser we used in the following conditions: 0.3 mm slits, Epass 100 eV, 0.1 eV / 1.176 s. The resolution of the analyser in these conditions is around 0.075 eV. Valence band spectra were acquired in 3 minutes whereas Ag $2p_{3/2}$ spectra were acquired in 12 minutes.

A standard silver deposit on a silicon wafer has been analysed after having been $Ar^+$-bombarded and annealed in vacuum.

<u>XAS.</u> XAS experiments were performed on the ROCK beamline of the SOLEIL synchrotron. Ag K-edge spectra were acquired during the single bunch mode of the machine at 16 mA, in order to have a minimal photon flux. In these conditions it is only $5 \times 10^9$ ph s$^{-1}$ around the Ag K-edge energy. The X-ray beam provided by a bending magnet is horizontally focused by a toroidal mirror; two mirrors then reject the high harmonics using the Pt stripe and vertically focalize the beam on the sample position. In order to lower the photon density on the sample, the beam is defocused to 0.86 mm². The two latter mirrors surround the Quick-EXAFS monochromator, which consist in a Si (2 2 0) channel cut monochromator mounted on an oscillating cam. We chose an acquisition frequency of 2 spectra per second. The chosen energy steps are 2 eV between 25005 eV and 25477 eV, 0.5 eV between 25478 eV and 25528 eV, 1 eV between 25529 eV and 25557 eV, and 2 eV between 25558 and 26678 eV. Spectra were extracted and averaged with the quickread.moulinex software and normalized with the Extranormal software, both were developed at the SOLEIL synchrotron, respectively by E. Fonda and C. La Fontaine, and O. Roudenko[26]. EXAFS signals and their FTs were processed with the ATHENA software[27]. Normalisation were made between 500 and 150 eV before the edge and between 42.5 eV and 1100 eV after the edge, with a background approximated by a 3rd order polynomial. EXAFS signals FTs were made in k$^3$, with a Hanning apodisation windows between 3 and 12 Å$^{-1}$.

Ultra dry silver chloride obtained from Alfa Aesar (Stock N°35715) has been used as a standard for XAS analysis. Pure pellets were pressed at 8 t for 30 s. For the sensitized layers, seven layers stacks were analysed, yielding an edge jump close to 1.

<u>UV-visible spectroscopy.</u> The UV-Visible spectroscopic measurements were performed on a CARY5000 spectrophotometer equipped with an integrating sphere in the total reflectance and transmittance modes (1 nm step, 1 s nm$^{-1}$). The absorptance was calculated as follows.

$$Absorbtance = 1 - Reflectance - Transmittance$$

AgCl for UV-visible spectroscopic measurements was precipitated from a 0.2 mol L$^{-1}$ AgNO$_3$ solution and a 1 mol L$^{-1}$ HCl solution, filtered, rinsed, dried and then pressed into a pellet.

<u>XPS.</u> Samples are excited by a monochromatised Al-Kα (1486.6 eV) X-ray source producing a 3.5 × 0.2 mm² beam. Photoelectrons are analysed by an hemispherical Phoibos 100 SPECS analyser in the following conditions: Epass: 20 eV, 0.1 eV / 0.1 s.

<u>XRF.</u> The Elio XGLab that was used is equipped with Rh source working at 40 keV, 100 mA, 480 s, producing a 1 mm beam and a SDD detector.

<u>XRD.</u> A Bruker D2 PHASER was used, in the following conditions: λ$_{Cu}$, 0.01° / 2 s.

**Equation S-1.** Dose formula for X-ray experiments.

$$Dose(\text{Gy}) = \frac{Photon\,flux(\text{ph.s}^{-1}) \times E_{photon}(\text{J}) \times \sin\alpha \times Exposure\,time(\text{s}) \times (1 - Transmittance)}{Beam\,surface(\text{m}^2) \times Thickness(\text{m}) \times d_{AgCl}(kg.\text{m}^{-3})}$$

**Equation S-2.** Dose formula for STEM experiments.

$$Dose(\text{Gy}) = \frac{Beam\,intensity(\text{A}) \times Exposure\,time(\text{s}) \times Total\,Stopping\,Power_{AgCl,E}(\text{J.m}^2.\text{kg}^{-1})}{Electron\,charge(\text{C}) \times Number\,of\,lines \times image\,length(\text{m}) \times Beam\,diameter(\text{m})}$$

**Equation S-3.** Dose formula for SEM experiments.

$$Dose(\text{Gy}) = \frac{Beam\,intensity(\text{A}) \times Exposure\,time(\text{s}) \times E_{elecctron}(\text{J})}{Electron\,charge(\text{C}) \times Scanned\,area(\text{m}^2) \times Absorbing\,thickness(\text{m}) \times d_{AgCl}(kg.\text{m}^{-3})}$$

**Equation S-4.** Formula for the calculation of the sensitized layer thickness, assuming all the silver is transformed in silver chloride. M represent a molar mass, ρ a density.

$$Sensitized\,layer\,thickness = Ag\,layer\,thickness \times \frac{\rho_{Ag} \times M_{AgCl}}{\rho_{AgCl} \times M_{Ag}}$$

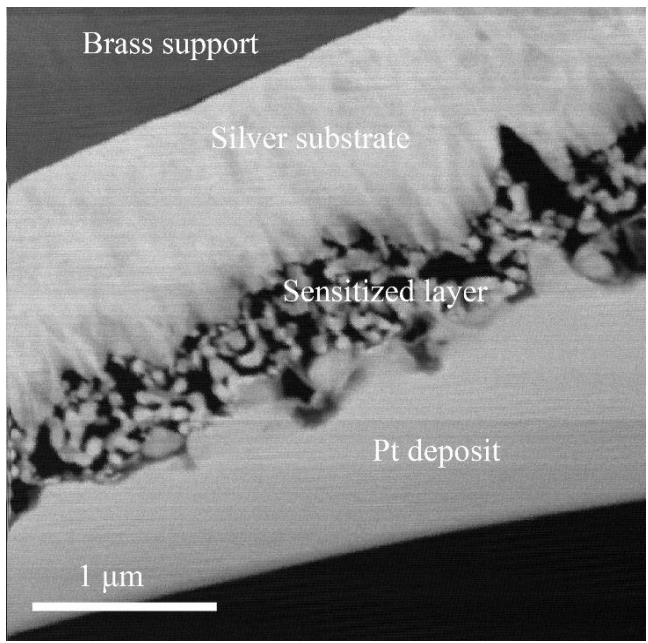

**Figure S-1.** STEM-HAADF image of a FIB-prepared sensitized real sample. The sensitized layer appears rough and porous; it has probably been damaged by the ion beam.

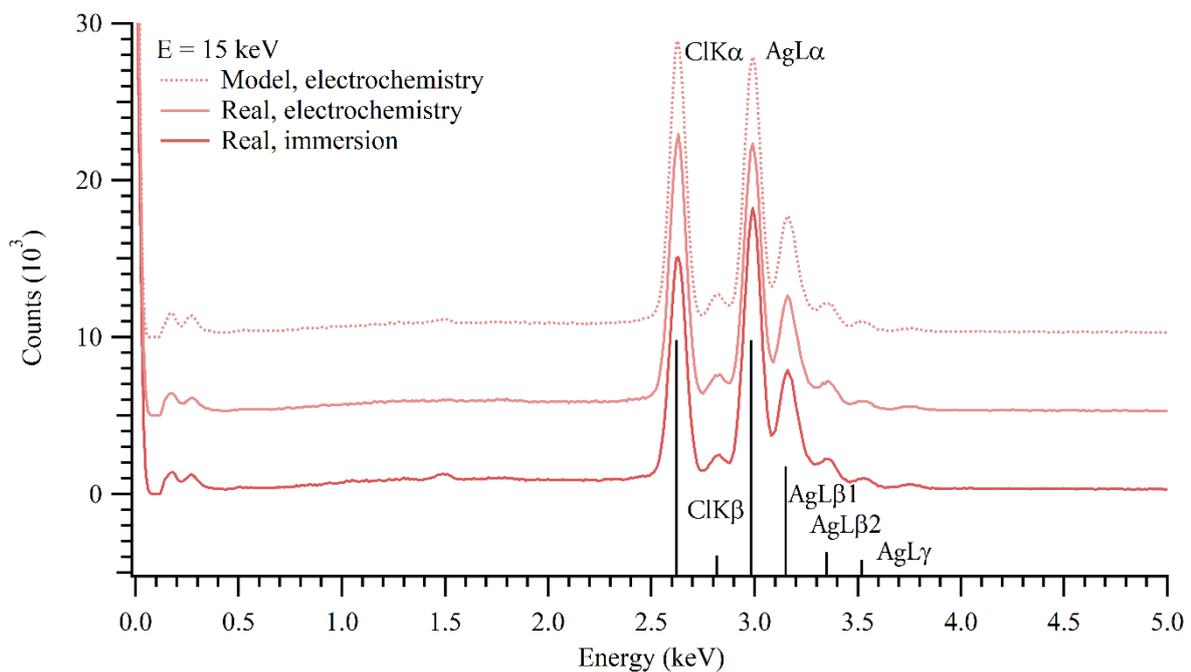

**Figure S-2.** EDX spectra of model and real samples sensitized by electrochemistry and immersion. The Ag/Cl ratio calculated by the φ(ρz) method are 0.93 for the electrochemically sensitized samples (model and real) and 1.17 for the immersion-sensitized real sample.

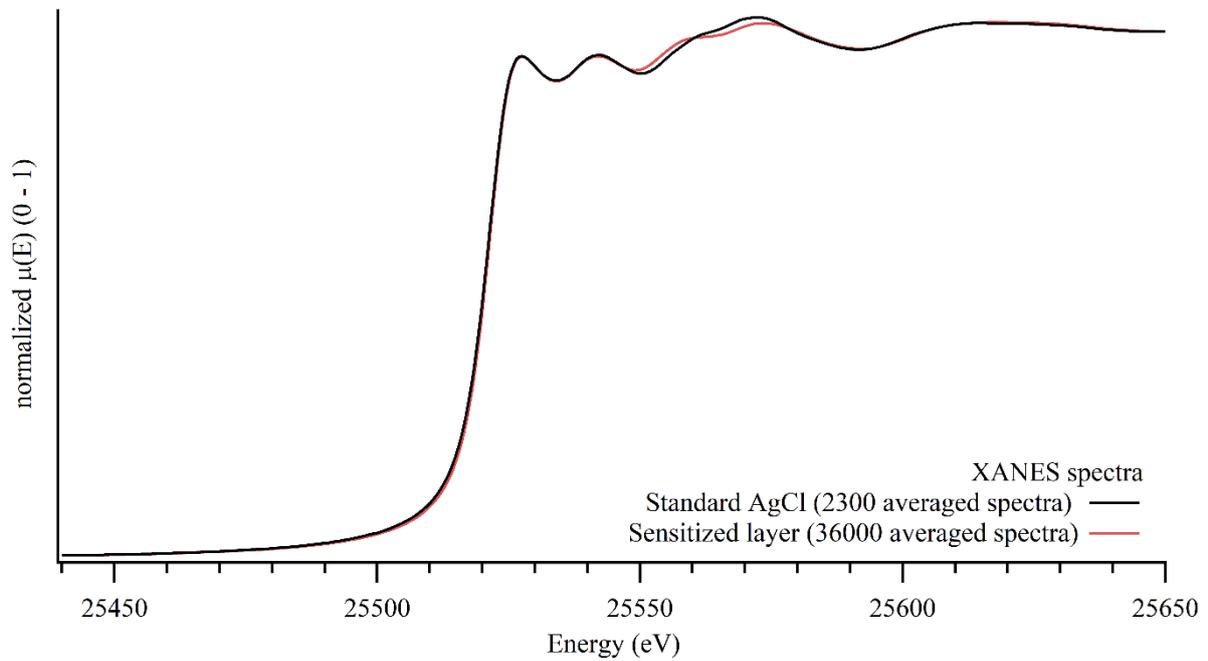

**Figure S-3.** XAS spectra of a sensitized layer and a standard AgCl pellet.

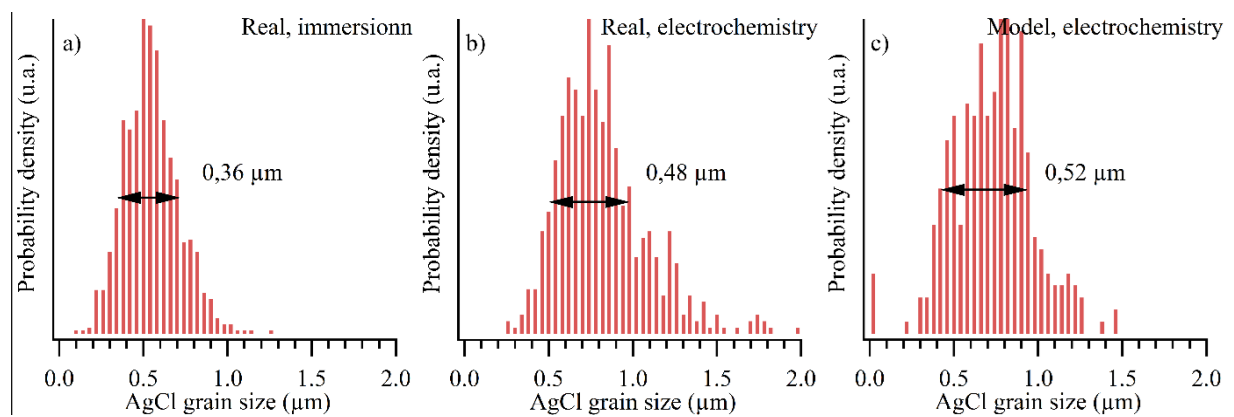

**Figure S-4.** Silver chloride grain size distributions, determined from plane view images obtained in the SEM, for real samples sensitized by immersion (a), by electrochemistry (b) and for model samples (c).

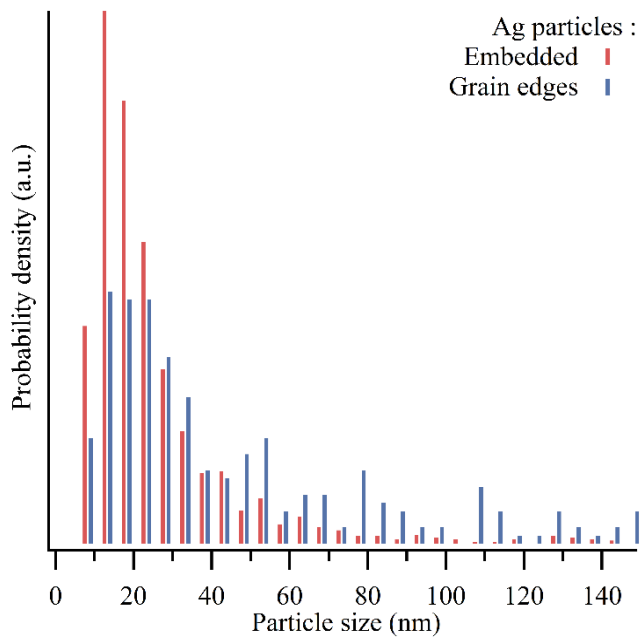

**Figure S-5.** Silver nanoparticles size distribution, according to their localization. 1451 embedded nanoparticles and 268 grain-edge nanoparticles were measured.

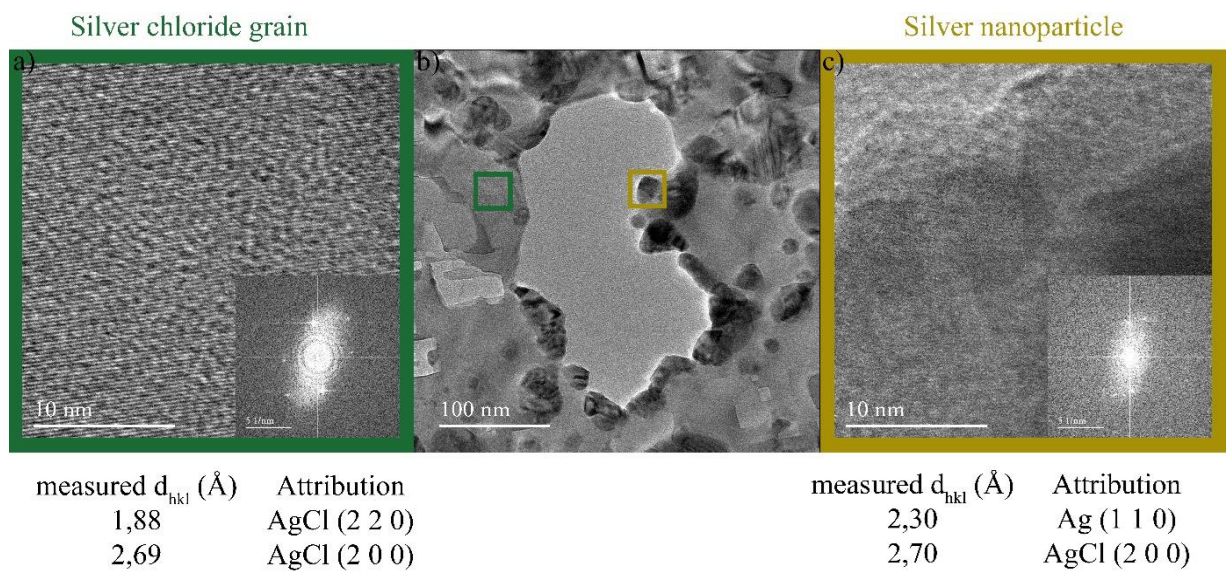

| measured $d_{hkl}$ (Å) | Attribution |
|---|---|
| 1,88 | AgCl (2 2 0) |
| 2,69 | AgCl (2 0 0) |

| measured $d_{hkl}$ (Å) | Attribution |
|---|---|
| 2,30 | Ag (1 1 0) |
| 2,70 | AgCl (2 0 0) |

**Figure S-6.** HRTEM images of a model sample. b) Global view. a) HRTEM image of the area circled in green, which corresponds to a silver chloride grain. Inset: FT of the HRTEM image. Table: $d_{hkl}$ measured on the FT and the plane attributions. c) HRTEM image of the area circled in yellow, which corresponds to a silver nanoparticle. Inset: FT of the HRTEM image. Table: $d_{hkl}$ measured on the FT and the plane attributions.

|  | R-factor | Sphere number | Crystallographical parameters | | $\Delta E_0$ (eV) | Fits results | | |
|---|---|---|---|---|---|---|---|---|
|  |  |  | Coordinence | Distance (Å) |  | Coordinence | Distance (Å) | $\sigma^2$ (Å²) |
| Standard AgCl | 0.0673 | 1 | 6 | 2.773 | -2.28 | 2.163 | -0.088 | 0.0127 |
|  |  | 2 | 12 | 3.922 |  | 3.059 | -0.124 | 0.0305 |
| Sensitized sample | 0.0251 | 1 | 6 | 2.773 | -0.257 | 2.163 | -0.0682 | 0.0171 |
|  |  | 2 | 12 | 3.922 |  | 3.059 | -0.0964 | 0.0343 |

**Table S-1.** Results of the fits of the standard AgCl pellet and of the sensitized sample between 1 and 5 Å (Hanning apodisation window [3; 10] Å$^{-1}$) including the first two coordination shells of AgCl. The S0² factor have been set to 0.78, according to the first fit made on the TF of the EXAFS signal of the sensitized samples, which has been acquired with a higher statistic.

**Experimental section.**

Materials. Real samples are made from 30 μm-silver coated 20 × 20 mm² brass platelets. These were ordered from the silversmith François Cadoret, 172-174 rue de Charonne – 75011 Paris, France (no longer in business). They are ¼ μm-diamond polished and cleaned in ethanol before the sensitization. Only silver was detected by XRF on the gilded side. The brass support contains traces of Cr and Co. Model samples are made from 1 μm thick silver foils, ordered from the silversmith Mario Berta Batilloro, Cannaregio 5182 – 30121 Venezia, Italia. Silver and traces of Fe, Cu, Zn and Cr have been detected by XRF in these foils.

SEM-FEG. SEM samples are protected with a carbon deposit made with a Leica EM SCD500 evaporator at 5×10$^{-4}$ mbar. The microscope is a SEM-FEG Zeiss Ultra 55 working at 1.5 kV, with a 10 μm aperture at a working distance ranging from 1 to 2 mm. The "high current" mode is not activated. 1024 × 768 pixels images are taken with the following parameters: "scan speed" 9 and "line average" 3 (1 min/image). The specific detection of SE1 is made with an in-column detector called "In Lens" on this type of microscope. This obliges to work at a low energy and offers a better spatial resolution, but lowers the depth of field. For EDX analysis with the SDD Bruker X-flash 4010 10 mm² detector, the microscope is operated in the following conditions: 15 kV, 120 μm aperture, 7.5 mm working distance, "high current", shaping time 90 kcps, energy range 40 keV, Cu Kα FWHM 146 eV. Spectra were acquired in 80 s. The approximated beam diameter onto the sample is 3 nm. The Optiprobe software allows the measurement of the beam intensity, which is 17 pA in the imaging conditions and 2798 pA in the EDX analysis conditions.

Size measurements of grains have been made with the Fiji software on 880 grains for real immersion-sensitized samples, 577 grains for real electrochemically sensitized samples and 326 grains on electrochemically sensitized model samples.

TEM-FEG. TEM samples were embedded in LR white resin, that polymerised at 50°C and a few 0.1 bar during 36 h. A Leica UC7 ultramicrotome was used, together with a Cryotrim diamond knife for the trimming and an Ultra 35 diamond knife for the cutting of ultrathin sections. The sections are cut with the layer sample parallel to the knife; the recto and the verso of the layers are identified before the embedding and at the end of the preparation, by comparison with the position of the sample in the resin section. 5 nm carbon deposits are made with a Leica ACE600 evaporator in the E-beam mode at $1\times10^{-5}$ mbar, at around 0.2 nm s$^{-1}$.

The microscope is a JEOL 2100F equipped with a Schottky FEG working at 200 kV, a JEOL STEM detector, a Gatan US4000 TEM camera, a JEOL Si(Li) 30 mm² EDX detector (138 eV resolution, bias: 700 V), and a cryo sample holder. The 50 µm aperture was used, with a 0.2 nm spot size for STEM-HAADF imaging. The 1024 × 1024 pixels images are acquired in 33.5 s. The same aperture was used for EDX analysis, with a 0.7 nm spot size. 512 × 512 pixels maps are acquired in ten minutes in the "T4" mode, favouring the energetic resolution over a high saturation limit. Lastly, the same aperture was used for HRTEM imaging, in the following conditions of brightness and beam convergence angle: "spot 1 / α3" for images at magnifications lower than 100 k and "spot 1 / α1" for higher magnifications. 4096 × 4096 images (binning 1) are acquired in 5 s. The phosphorescent screen picoamperemeter was used in order to determine the beam current, which is 57 pA for STEM imaging, 650 pA for EDX analysis and 1100 pA for TEM imaging.

Size measurements of nanoparticles have been made with the Fiji software on 1451 particles for embedded nanoparticles and 268 grain-edge nanoparticles. Thickness measurements have been made on 11 layers.

HAXPES. HAXPES experiments were performed at the GALAXIES beamline at the SOLEIL synchrotron. The X-ray beam provided by an undulator is monochromatised by a double-crystal Si(111) monochromator (DCM) and focused onto the sample position by a toroidal mirror located downstream. Incident energies of 10 keV or 9.9 keV X-ray were used. The third harmonic of the DCM was used in order to lower the photon flux and increase the energetic resolution. In addition, a 5.8 % transmitting filter was used to further attenuate the beam. The photon flux in these conditions was measured at $7 \times 10^9$ ph s$^{-1}$. The resolution of the monochromator at this energy is around 0.3 eV. The beam size was $20 \times 80$ µm² and the angle between the beam and the sample surface was 20°. The sample was cooled down at 200 K. The beamline is equipped with a EW4000 VGScienta hemispherical analyser we used in the following conditions: 0.3 mm slits, Epass 100 eV, 0.1 eV / 1.176 s. The resolution of the analyser in these conditions is around 0.075 eV. Valence band spectra were acquired in 3 minutes whereas Ag $2p_{3/2}$ spectra were acquired in 12 minutes.

A standard silver deposit on a silicon wafer has been analysed after having been Ar$^+$-bombarded and annealed in vacuum.

XAS. XAS experiments were performed on the ROCK beamline of the SOLEIL synchrotron. Ag K-edge spectra were acquired during the single bunch mode of the machine at 16 mA, in order to have a minimal photon flux. In these conditions it is only $5 \times 10^9$ ph s$^{-1}$ around the Ag K-edge energy. The X-ray beam provided by a bending magnet is horizontally focused by a toroidal mirror; two mirrors then reject the high harmonics using the Pt stripe and vertically focalize the beam on the sample position. In order to lower the photon density on the sample, the beam is defocused to 0.86 mm². The two latter mirrors surround the Quick-EXAFS monochromator, which consist in a Si (2 2 0) channel cut monochromator mounted on an oscillating cam. We chose an acquisition frequency of 2 spectra per second. The chosen energy steps are 2 eV between 25005 eV and 25477 eV, 0.5 eV between 25478 eV and 25528 eV, 1 eV between 25529 eV and 25557 eV, and 2 eV between 25558 and 26678 eV. Spectra were extracted and averaged with the quickread.moulinex software and normalized with the Extranormal software, both were developed at the SOLEIL synchrotron, respectively by E. Fonda and C. La Fontaine, and O. Roudenko[25]. EXAFS signals and their FTs were processed with the ATHENA software[26]. Normalisation were made between 500 and 150 eV before the edge and between 42.5 eV and 1100 eV after the edge, with a background approximated by a 3rd order polynomial. EXAFS signals FTs were made in k$^3$, with a Hanning apodisation windows between 3 and 12 Å$^{-1}$.

Ultra dry silver chloride obtained from Alfa Aesar (Stock N°35715) has been used as a standard for XAS analysis. Pure pellets were pressed at 8 t for 30 s. For the sensitized layers, seven layers stacks were analysed, yielding an edge jump close to 1.

<u>UV-visible spectroscopy.</u> The UV-Visible spectroscopic measurements were performed on a CARY5000 spectrophotometer equipped with an integrating sphere in the total reflectance and transmittance modes (1 nm step, 1 s nm$^{-1}$). The absorptance was calculated as follows.

$$Absorbtance = 1 - Reflectance - Transmittance$$

AgCl for UV-visible spectroscopic measurements was precipitated from a 0.2 mol L$^{-1}$ AgNO$_3$ solution and a 1 mol L$^{-1}$ HCl solution, filtered, rinsed, dried and then pressed into a pellet.

<u>XPS.</u> Samples are excited by a monochromatised Al-Kα (1486.6 eV) X-ray source producing a 3.5 × 0.2 mm² beam. Photoelectrons are analysed by an hemispherical Phoibos 100 SPECS analyser in the following conditions: Epass: 20 eV, 0.1 eV / 0.1 s.

<u>XRF.</u> The Elio XGLab that was used is equipped with Rh source working at 40 keV, 100 mA, 480 s, producing a 1 mm beam and a SDD detector.

<u>XRD.</u> A Bruker D2 PHASER was used, in the following conditions: λ$_{Cu}$, 0.01° / 2 s.